\documentstyle[12pt]{article}
\begin{document}

\baselineskip8mm
\def\be{\begin{equation}}
\def\ee{\end{equation}}
\def\ni{\noindent}

\sloppy
\title {Universality classes for the "ricepile" model with absorbing properties}
\vskip 1cm
\author{M\'aria Marko\v sov\'a\\
Dept. of Computer Science and Engineering,\\
 Faculty of Electrical Engineering
and Information Technology\\
Slovak University of Technology\\
Ilkovi\v cova 3, Bratislava, Slovakia}

\maketitle
\begin{abstract}
The absorbing "ricepile" model with stochastic toppling rules has been numerically 
studied. Local limited, local unlimited, nonlocal limited and nonlocal unlimited 
versions of the absorbing model have been investigated. Transport properties and different dynamical 
regimes of all of the models have been analysed, from the point of view of self 
organized criticality (SOC). Phase transitions between different dynamical regimes 
were studied in detail. It was shown, that the absorbing models belong to two 
different universality classes.
\end{abstract}
\newpage
\section {Introduction}

Self organized criticality (SOC) is widely studied phenomenon in the last ten years. Theory of SOC, proposed by Bak, 
Tang and Wiesenfeld 
\cite{BTW}
describes dynamical behaviour of many particle systems with local interactions. 
Paradigmaticaly, the description is based on the dynamics of a pile of sand.  If the sandpile is randomly driven
by  slow addition of sandgrains, the slope of pile grows up and after some time 
local stability conditions are violated somewhere on the pile 
surface. The avalanche starts to slide down the slope. This type of dynamics is easily 
modeled by a cellular automaton. Such a model "sandpile" is defined on a large n- dimensional lattice.  
The avalanche dynamics is, under the action of slow drive, governed by the local critical 
conditions (such as the local critical slope, for example), and local toppling rules.
This dynamics leads to 
the steady state called self organized critical state, characterized by the critical scaling of 
the avalanche size distribution:
\be\label{p(s)}
p(s,L)=s^{-\tau} f({s\over {L^{D}}})
\ee

In 
(\ref{p(s)}),  $L$ is the system size and $s$ exhibits the  size of the avalanche.
Critical exponents $\tau$ and  $D$, depend on the particular  model. 
Different models can be divided 
into  the various universality classes, defined by a specific set of the critical exponents
\cite{KAD,Al3}
.

A natural step from the sandpile model systems, leads to the investigation of real 
piles of granular
material from the point of view of SOC dynamics. Several efforts have been made  
in this direction
\cite{Nag,ER,HG},
but no clear evidence of self organized critical dynamics (\ref{p(s)}) has been found.
Finaly, in 1996 an experiment has been done by the group of experimentalists and theoretists in Oslo
\cite{Oslo}.
 In the Oslo experiment dynamical behaviour of the driven quasionedimensional pile of rice has been 
investigated. The avalanche sizes in the steady state were measured in terms of dissipated potential energy.
Two type of grains were used (elongated and round ones), showing completely different dynamics. In the 
case of the ricepile, consisting of elongated grains, SOC state has been  established, in which 
the avalanche size 
distribution has a power law character with critical exponent $\tau \approx 2.02$
\cite{Oslo}
.
   
Soon after the experimental results were published, the  model "ricepile" cellular automata 
were suggested and numerically studied
\cite{Al1,Al2,Al3,CHC}
.The model ricepile is, in principle, cellular automaton defined on onedimensional lattice, with 
randomness incorporated into the toppling rules and with deterministic drive. Changes in toppling
rules are often manifested by  different dynamical behaviour of the model and different 
universality class into which the 
model belongs. 
\cite{CHC, KAD}

In this paper we study a model, which exhibits a modification of the two threshold ricepile model
\cite{Al1,Al2}. In \cite{Al1,Al2}  the gravity effects, grain friction and the local conditions on the pile
are described by the parameter $p$. Two thresholds, namely the  critical threshold
and the gravity threshold are defined. Thresholds governs the movement of the grain on the pile 
surface. 
 
We removed  the second, gravity  threshold of the ricepile model. This causes, that the model, 
depending on the parameter  $p$
value, has absorbing properties and 
interesting dynamical regimes
\cite{Mark}. 
One threshold model in two dimensions has been studied by Tadi\' c and Dhar
\cite{TDh}.
 
Here, different versions of the absorbing model are defined using  different toppling rules. Dynamical 
behaviour of the local limited ({\bf LLIM}), local unlimited ({\bf LUNLIM}), nonlocal limited ({\bf NLIM}) and nonlocal
unlimited ({\bf NUNLIM}) absorbing  model is studied and two distinct universality classes are recognized.

\section{Ricepile models}

The experimental results of the Oslo group motivated further theoretical studies of the avalanches and dynamics 
of granular material
\cite{PB,HR}
. The main question is, which  physical properties of the pile granular material, are important for 
SOC state to be established. What is, for example, the role of friction, what is the role of the grain shape,   
the gravity and inertia of rolling grains? 
The ricepile experiment reveals, that no SOC is possible if the ricepile 
consists of round ricegrains.  On the contrary, the dynamics of pile consisting of elongated grains, is self 
organized critical
\cite{Oslo}
. Certainly, the shape of grains, and the additional  effects related to  the shape, such  
as better packing of  grains due to the elongated shape, supressed 
rolling of grains and thus supressed inertia effects, are of great importance
\cite{Oslo,HR}.

Ricepile models are  cellular automata, in which friction and gravity effects are taken into 
account in a simple way, through the parameter  $p$.
The value of the parameter $p$ decides, whether the grain will stop on the site, or roll
further down the slope. 

The two threshold ricepile model introduced in 
\cite{Al1, Al2}
is defined on a onedimensional lattice of size $L$, with a wall at the zero position and open boundary at the other end.
At the open boundary, particles are free to flow out of the system.

As in the experimental set up, the system is driven by adding 
 particles to the position one, at the closed end. Every time unit one particle is added. 
Two thresholds are defined in the model: the critical threshold $z_c$, which is the local condition 
for the onset of avalanche, and the gravity threshold $z_g$ ($z_c < z_g$). If the local slope
\be
\label{z}
z_i=h_i-h_{i+1},
\ee
(where $h_i$, $i=1,2,3,...,L$ is a height profile of pile) is less then $z_c$, it is too
low and the friction stops the grain movement. The grain resides on the position $i$. If $z_i > z_g$,
the local slope is too high and the grain moving downslope is not allowed to stop at the i-th 
position. But, in the  case, that $z_c < z_i < z_g$, the grain moves from $i$ to $i+1$ with probability
$p$. Through the parameter $p$, effective friction is introduced into the model. All supercritical 
slopes can topple, with probability $p$, but for local slopes, which are too big  $(z_i > z_g)$, 
the gravity 
becomes decisive and the site topples with probability  $p=1.0$. 

The dynamics of the ricepile model 
\cite{Al1,Al2} 
is as follows:

\medskip

\ni 1. Each avalanche starts at $i=1$. If $z_1 > z_c$,  the site one is activated and topples 
a particle to the 
next nearest position (two), with the probability $p$. If even $z_1 > z_g$;  $p=1.0$.

\medskip

\ni 2. Every particle, sliding from the position $i$ to $i+1$ activates three columns, namely $i-1$, $i$ and 
$i+1$. The position $i-1$ is activated, because it possibly can become supercritical , when removing a grain
from the $i$ - th column. Columns $i$ and $i+1$ are activated, 
because they are destabilized by sliding or stopping particles, respectively. In the next time step, all
supercritical active sites  
topple a particle to the $i+1$-st position with probability $p$ 
($p=1.0$,  if $z_i > z_g$).

\medskip

\ni 3. Step two is repeated, untill there are no active sites in the system, that means, 
untill the avalanche is not over.

\medskip

Changing slightly the local toppling rules, different versions of the ricepile model are defined
\cite{Al3}
:

\medskip

\ni a) The number of particles toppled from the position $i$ is constant  and independent on 
the supercritical local slope $z_i$ - the model is 
called {\bf limited}.

\medskip

\ni b) The number of toppled particles is a function of  the supercritical local slope $z_i$, the model
is called {\bf unlimited}.

\medskip

\ni c) If the particle (or more particles) topples from the site $i$ and moves only to the next 
nearest position $i+1$, the model is defined as {\bf local}.

\medskip

\ni d) The model is called {\bf nonlocal}, if $n$ toppled particles moving from the $i$-th site
are added subsquently to $n$ nearest downslope positions (one particle per site) $i+1$, $i+2$,..., $i+n$.

\medskip

\ni Thus four different ricepile models are recognized:

\ni 1. local limited model ({\bf LLIM})

\ni 2. local unlimited model ({\bf LUNLIM})

\ni 3. nonlocal limited model ({\bf NLIM})

\ni 4. nonlocal unlimited model ({\bf NUNLIM})

Universality classes for the ricepile model were studied by Amaral and Lauritsen 
\cite{Al3}
. Their results show, that local models (LLIM, LUNLIM) belong to the wide universality class called 
local linear interface universality class (LLI class)
\cite{Al3,PB}
. 
The authors also found, that the nonlocal  toppling rules lead to two new  universality classes, 
with a different set of critical exponents. 
But none of the universality classes is the one of the real ricepile.

\section{Absorbing model}

Our  absorbing model 
\cite{Mark}
exhibits simplified, one threshold version of the ricepile model
\cite{Al1}
. The gravity threshold is removed and all supercritical active sites are alloved  to 
topple with probability $p < 1$. There , therefore, persists small, but nonzero probability, that also the 
extremely  
large local slopes are possible. Physically this seems to be quite plausible.
It is not probable, that in real piles of granular material there exists a strict 
gravity threshold. There is rather a continuous transition to the local slopes ,
which are already so large that they, when activated, {\bf almost} always topple. 

We investigated numerically all four versions of the absorbing model: LLIM, LUNLIM, NLIM, NUNLIM. 
Several quantities were measured for all of the models:

\medskip

\ni 1. Material transport as a ratio of the number of outgoing to ingoing particles
\be\label{transport}
J(p)={n_{out}(p)\over n_{in}(p)},
\ee
and its dependence on the parameter $p$.

\medskip

\ni 2. Average material transport $\langle J(p) \rangle$ as a function of $p$.

\medskip

\ni 3. Avalanche size distribution for different parameter values $p$. Avalanche sizes are 
measured in terms od dissipated 
potential energy, in accordance with the experiment
\cite{Oslo}.

\medskip

\ni 4. Changes in the pile profile, due to changes in the parameter $p$ value.

\medskip

If the probability  parameter $p$ changes slowly in the interval $(0,1)$,
the model typically passes through different dynamical regimes: 

\ni i) isolating, in which all particles
are absorbed in the system and none of them  
reaches the open boundary; 

\ni ii) partialy conductive, in which the pile 
profile grows up as a bulk, because a certain fraction of the particles,
depending on $p$,  is absorbed in the system 
({\bf absorbing properties});
 
\ni iii) and totaly conductive, when the number of ingoing and outgoing particles 
is balanced.

\section{Universality classes for the absorbing  model}

\subsection{LLIM}

Dynamical properties of the local limited absorbing model are in details described in
\cite{Mark}
. Here I only briefly list the main results.

Local limited toppling rules are defined as follows:  active supercritical site topples 
one particle to the next nearest position with the probability $p$. 
Looking at the average material transport $\langle J(p) \rangle$, three dynamical regimes of the LLIM model are 
recognized $(Fig.1a)$:

\medskip

\ni{\bf a)} For $0 < p <p^{'}$; ${p^{'}}\approx 0.53865$, the system is completely isolating. 
The average transport 
$\langle J(p) \rangle$ is zero. For $p$ close to zero, almost all ingoing particles are absorbed. The avalanches die out 
soon , their size is exponentially bounded. 

Close to the first phase transition point $p^{'}$, the steepness of the pile is still high enough to say, 
that the local slopes are almost everywhere higher than the critical threshold $z_c$. This is the reason 
that the spreading of active sites in time is practically determined by the probability $p$; the same way
as it is in the percolation process.
In the space - time coordinate system, we have therefore a picture of directed percolation with three 
descendants and absorbing boundary
\cite{LSMJ}
.  $p^{'}$ is thus simply the critical percolation threshold. Close to the percolation 
threshold $p^{'}$, the average
transport $J(p)$ scales with $p$ as
\be\label{1st}
{{\langle J(p) \rangle-J^{'}}\propto {(p-p^{'})^{\delta^{'}}}}
\ee
$$\delta^{'}= {0.9\pm 0.01}$$

\ni where $J^{'}$ is the current flowing due to the finite size of the system.

\medskip

\ni{\bf b)} For the probability interval ${p^{'}} < p < p_c$; $p_c\approx 0.7185$, the system is partialy 
conductive, with constant average slope. That means, the height profile grows as a bulk with velocity
$v(p)$. Fluctuations of transport $J(p)$ exhibit long range correlations. 

Above the percolation threshold $p^{'}$, the percolation picture breaks down. The subcritical, 
absorbing states
are randomly distributed throughout the system and the avalanche can stop anywhere. As $p\to p_c$, the 
long range correlations in transport fluctuations are destroyed, and the region of small local slopes 
spans the whole system. The pile stops to grow and at $p=p_c$ it is pinned at the position $i=L$.  Critical 
point $p_c$ is thus 
understood as the depinning transition point. Close to the depinning critical point, average transport
scales as

\be\label{sek}
1-\langle J(p)\rangle\propto {\mid p-p_c \mid^{\delta}}
\ee
$$\delta= {0.9\pm 0.01}$$

\ni{\bf c)} In the interval $(p_c,1)$   the system is completely conductive. Transport fluctuations 
are of white noise type and the average transport $J(p)=1.0$.
In the dynamical regimes b) and c) the system is in the SOC state, having power law distribution of avalanche 
sizes 
(\ref{p(s)}) with critical exponent $\tau = 1.57\pm 0.05$. 

\subsection{LUNLIM}

Local unlimited toppling rules are in the absorbing  model defined as follows: In order to get a realistic 
profile of the pile, each supercritical active site topples $k$, $k=int({z_i\over 2.0})$, 
grains to the next nearest downslope position with probability $p$.
This way one gets smooth profile without cavities $(Fig.2a)$.

Numerical investigations of $\langle J(p)\rangle$ reveals, that only two dynamical regimes are clearly recognized 
$(Fig.1b)$: the pile is either completely isolating, $(\langle J(p) \rangle = 0)$, or completely 
conductive $(\langle J(p)\rangle = 1.0)$.
Partialy conductive dynamical regime is missing. 

\ni{\bf a)} For $0 < p <p_c$; $p_c\approx 0.6995$ $(Fig.1b)$, the system is completely isolating.
From the definition of local toppling rules it is clear, that absorbing states $(z_i < z_c)$ are easily
created even for very small values of the parameter $p$. 
 It means, that the percolation picture in the space - time coordinate 
system is not correct in the  case of the LUNLIM model. Near transition point $p_c$ the average 
transport $J(p)$
scales with $p$ as $(Fig.3)$
\be\label{jj}
{J(p)\propto (p-p_c)^{\delta}}
\ee
$$\delta = 1.93\pm 0.07$$

\ni {\bf b)} In the second dynamical regime $(p_c < p < 1.0)$ the system is completely conductive, with 
pile profile pinned at $i=L$ $(Fig.2a)$. $J(p)$ as a function of time exhibits white noise features $(Fig.4)$.
Avalanche size distribution is critical
(\ref{p(s)})
with critical power law exponent $\tau = 1.54\pm 0.02$ $(Fig.5a)$.

\subsection{NLIM}

Nonlocal limited toppling rule means, that the supercritical active site topple, with probability 
$p$, $N$ particles 
to the $N$ nearest downslope positions.
The nonlocal limited toppling rules preserve three dynamical regimes, the same way as it is in the LLIM case.
In $Fig.1c$,  isolating, partialy conductive and totaly conductive regimes are recognized.

\ni {\bf a)} For $0 < p < p^{'}$, ${p^{'}}\approx 0.267$ $(Fig.1c)$, the pile is in the 
isolating regime. 
To understand 
the nature of the first phase transition point $p^{'}$, the percolation picture in the time - space 
coordinate system is still usefull. But now, the number of descendants  is, in principle, greater 
 than three.  That is the reason for the fact, that the percolation threshold is shifted 
to the lower parameter values as one can see when comparing $Fig.1a$ and $Fig.1c$, e. g. 
$({p^{'}}_{\rm NLIM} < {p^{'}}_{\rm LLIM})$. In the model studied here, the number of particles toppling 
from the activated supercritical site is $4$. Five  sites are thus activated by every toppling from 
the position $i$, namely $i-1$, $i$, $i+1$, $i+2$, $i+3$. There are therefore five descendant  sites
in the directed percolation with absorbing boundary
\cite{Mark}
. In order to estimate $p^{'}$ with greater accuracy, systematic studies of the dependence of percolation 
threshold on the number of descendant sites are necessary.

 The average transport near the percolation threshold $p^{'}$ scales as $(Fig.6a)$:
\be
{J(p)\propto (p-p^{'})^{\delta^{'}}}
\ee
\ni with the critical exponent

$${\delta^{'}}={1.18\pm 0.04}$$

\medskip

\ni {\bf b)} In the interval ${p^{'}} < p < p_c$, the pile grows up with constant velocity $v(p)$, 
maintaining 
the global slope on a constant value for a constant probability parameter $p$. Transport $J(p)$ as a
function of time shows long range correlations, on the contrary to the totaly conductive
regime, where it has a character of white noise $(Fig.7)$.

\ni {\bf c)} Depinning transition occurs at $p_c \approx 0.365$ $(Fig.6b)$. Average transport scales with 
$p$ close to the critical point as :
\be
1-\langle J(p)\rangle \propto \mid p-p_c \mid^{\delta}
\ee

$${\delta = 1.12\pm 0.06}$$

For the probability interval
$p_c < p < 1$, the profile of pile is pinned at $i=L$ $(Fig.2b)$, and the average transport $\langle J(p)\rangle = 1.0$ 
$(Fig.1c)$. $Fig.5b$ demonstrates the avalanche size  distribution in a case of partialy conductive 
and conductive dynamical regimes. In both cases the  dynamics of the pile is self organized critical,
which is demonstrated by the critical , power law scaling
(\ref{p(s)}). 
The  critical exponent $\tau = 1.35 \pm 0.05$.
  
\subsection{NUNLIM}

The nonlocal unlimited toppling rule is defined as follows: $N(z_i)$ particles are released 
(with probability $p$)
from the activated supercritical
position and are added to $N(z_i)$ nearest downslope positions.
 
The nonlocal unlimited version of the absorbing model shows completely different behaviour. First,
no distinct dynamical regimes are recognized. The pile is completely conductive already
for $p$ close to zero as can be  seen from $Fig.8a$. Pile profile is pinned at $i=L$ $(Fig.2c)$. 
Avalanche size distribution shows the critical scaling 
(\ref{p(s)}) 
with
critical exponent $\tau = 1.51 \pm 0.05$ $(Fig.5e)$. 

\section{Discussion and conclusion}

Probability density function 
(\ref{p(s)})
scales with the system size as 
\be
p(s,L)=L^{-\beta} g({s\over {L^D}})
\ee

\ni with $\beta = D \tau$. For the LLIM, LUNLIM and NUNLIM absorbing models the best data collapse
has been found for $D=2.24$, what indicats, that these models belong to the same universality 
class, called LLI universality class 
\cite{Al3, PB}
.

On the contrary, for the NLIM absorbing model the best data collapse was found for $D=1.55$. 
The critical exponents 
$\tau$ and $D$ are different from that of LLI class and defines a new universality class to which belongs 
also the NLIM version of the two threshold ricepile model 
\cite{Al3}
.

The reason of lowering  the $\tau$ exponent of the NLIM model in comparison with the LLIM model is as follows:
The average slope of the NLIM and the LLIM pile is simillar. For example for $p=0.8$, the average slope 
of the NLIM pile is $67.87^o$ and for the LLIM pile $60.53^o$. Due to the nonlocal toppling rules in the NLIM
model, more columns
are perturbed and thus the probability of greater avalanches is enhanced. Therefore the 
exponent  $\tau$ is lowered.

The same argument could be used in the case of the NUNLIM and the LUNLIM model. 
But here the situation is different.
The average slope changes significantly with changes in toppling rules from local to nonlocal. For example 
if $p=0.8$, the average slope of the LUNLIM model is $82.22^o$ and that of the NUNLIM model 
equals $53.6^o$. Because the number 
of toppled particles is proportional to the slope in the unlimited model, it seems, that the relatively small
average slope of NUNLIM pile leads to relatively few particles released on average in one toppling. 
This fact should enhance the probability of small avalanches and the avalanche size distribution  
function should have 
$\tau$ exponent greater than $1.55$. This really happens for the NUNLIM two threshold ricepile model
\cite{Al3}
. In this model the propability of big local slope decreases exponentially with $z_i$.
But looking at $(Fig.2c)$ one can see, that it is not an exception to have a big local slopes in   the 
NUNLIM absorbing pile.
During a toppling event, the site with big local slope releases a number of particles 
(proportional to the local
slope), which disturb a lot of downslope columns.  This efect increases the probability
of large avalanches. It seems, that in the case of the absorbing model the two described effects 
balance each other
and thus the exponent $\tau$ remains untouched by changed toppling rules. This is different from the NUNLIM
two threshold model 
\cite{Al3}. 
Here the first effect is decisive and  the model belongs to a new universality class $(\tau = 1.63)$.

Another question, which should be discussed, is the nonexistence of partialy conductive regime of 
the LUNLIM model. First, the model is local. That means, in every toppling, only three columns are 
activated by each toppled particle. Therefore, the probability of avalanches, having a chance to 
reache the end of the system and thus to transport a material,  doesn't increase
due to more activated sites by every toppled particle. 
 As it has been already told, in the LLIM model, there are no absorbing $(z_i < z_c)$
states in the system for the isolating dynamical regime. This is not the case of the LUNLIM model.
Absorbing states, therefore, 
 exhibits another obstacle for bigger avalanches to develop and transport 
the particles. The existence of absorbing states, even for a small 
parameter values, destroys the percolation  picture of the spreading of active sites and this is 
also the reason for nonexisting critical percolation probability $p^{'}$ and only the transition 
to completely conductive dynamical regime is present.

Last, some words should be told about the pile slopes in all of the three dynamical regimes.
In the isolating regime average transport $\langle J(p) \rangle =0$. 
All the added particles are absorbed in the system.
Moreover, the avalanche sizes in this regime are exponentialy bounded. That means, majority of the particles is absorbed 
on the first few columns of the pile. Therefore, if the driving time $t$ tends to infinity, 
the average slope of the pile grows to infinity. That in consequence means, that also the local slopes
become arbitrarily large.

In the partialy conductive regime, constant amount of particles, depending on the parameter $p$, 
is absorbed in the system. As $t \to \infty$, average slope of the pile remains constant; pile grows up
as a bulk. The local slopes are finite, except of $z(L)$ (see $(2)$), which tends to infinity.

In the conductive regime $\langle J(p) \rangle =1.0$. The average slope of the pile is constant depending only 
on the parameter $p$. All local slopes are finite in this regime.

In conclusion, we have studied numerically the LLIM, LUNLIM, NLIM and the NUNLIM absorbing models. We have found, 
that the models 
belong to two different universality classes, characterized by different critical exponents.  
Both of the universality classes are different from the one of real pile of rice.  We have studied 
the transport properties of all of the models and found phase transitions between different dynamical 
regimes.
We state, that the dynamics of LLIM and NLIM model is directly mapped to the dynamics of directed percolation process
at the absorbing boundary 
\cite{LSMJ}
for the defined interval of parameter $p$. 

This work has been supported by the grant of VEGA number 2/6018/99. 

\newpage
{

}
\newpage

\section{Figure captions}

Fig.1

Average transport ratio of particles through the system as a funcion of  the parameter $p$.

\ni $(a)$ LLIM model: Three different dynamical regimes are recognized: isolating, $p \in (0,0.53865)$;
  partialy conductive, $p \in \langle 0.53865,0.7185)$ and conductive, $p \in \langle 0.7185,1.0\rangle $.

\ni $(b)$ LUNLIM model: Only two different dynamical regimes are recognized: isolating, $p \in (0,0.6995)$
  and conductive, $p\in \langle 0.6995,1.0\rangle $.

\ni $(c)$ NLIM model: Again, three different dynamical regimes are depicted: isolating,  $p \in (0,0.267)$;
  partialy conductive,  $p \in \langle 0.267,0.365)$  and totaly conductive, $p\in \langle 0.365, 1.0\rangle$.

\medskip

\ni Fig.2

Pile profiles of the LUNLIM $(Fig.2a)$, NLIM $(Fig.2b)$ and NUNLIM $(Fig.2c)$ absorbing models for different 
values of the probability parameter $p$.
Notice, that in the totaly conductive regime, the pile profile is pinned at $i=L$ $(a)$, $(c)$.
The pile is growing as a bulk with velocity $v(p)$ in  the case of the partialy conductive regime and is 
pinned in the totaly conductive regime $(b)$.  

\medskip

\ni Fig.3

Ln-ln plot of the average transport as a function of the distance from the critical point $p_c$ in 
the LUNLIM absorbing
model, $\epsilon = (p-p_c)$. We find, that the best scaling is obtained for $p_c=0.6995$.

\medskip

\ni Fig.4

Transport $J(p)$ as a function of time (in iterations)  in the conductive regime of the LUNLIM model for two 
different values of 
$p$. For $p \geq p_c$,  $p_c=0.6995$,
white noise is observed. For $p < p_c$ all particles are absorbed and therefore there are no 
fluctuations. 

\medskip

\ni Fig.5

Ln -ln plots of the power law  parts of  the avalanche size distributions (unnormalized). The critical 
exponent $\tau$ of the LLIM
\cite{Mark}, LUNLIM $(a)$ and NUNLIM $(c)$
models is $\tau =1.55$. The NLIM model $(b)$ belongs to the different universality class with  
$\tau = 1.35$. The system size in $Figs.5b,c$ is $L=300$ and in $Fig.5a$ $L=500$. In the NLIM 
model $(b)$ the bump shift with system size has been numericaly tested. With growing system size,
the bump shifts to the higher values of $s$, indicating thus SOC.

\medskip 

\ni Fig.6

\ni $(a)$ Ln-ln plot of the average transport as a function of the distance from the critical 
point $p^{'}$ in the NLIM absorbing
model, $\epsilon = (p-p_c)$. We find, that the best scaling is obtained for $p_c=0.276$.

\ni $(b)$ Ln-ln plot of the average transport as a function of the distance from the critical 
point $p_c$ in the NLIM absorbing
model, $\epsilon = (p_c-p)$. We find, that the best scaling is obtained for $p_c=0.3441$.

\medskip

\ni Fig.7

Transport $J(p)$ as a function of time (in iterations) in the partialy conductive regime $(p=0.339)$ and totaly conductive regime 
$(p=0.4)$ of the NLIM absorbing model.
For $p\geq p_c$,  $p_c=0.365$,
white noise is observed. For $p^{'} < p < p_c$ the character of fluctuations is different. Long range correlations 
(reminiscent of Brownian motion) are observable in the time signal.
For $p < p^{'}$  
all particles are absorbed and therefore there are no
fluctuations. 

\medskip

\ni Fig.8

Transport $J(p)$ as a function of time (in iterations) in the NUNLIM model for three different values of the 
parameter $p$. The pile is totaly 
 conductive in wide range of the parameter $p$ and the signal has a character of white noise, 
with fluctuations depending 
on $p$.


\begin{thebibliography}{99}
\itemsep0pt
\bibitem{BTW}
P. Bak, Ch. Tang, K. Wiesenfeld; Phys. Rev. Lett. {\bf 59}, 381 (1987)
%
\bibitem{Nag}
S.R. Nagel; Mod. Phys. {\bf 64}, 321 (1992)
%
\bibitem{ER}
P. Evesque, J. Rajchenbach; Phys. Rev. Lett. {\bf 62}, 44 (1989)
%
\bibitem{HG}
G.A.Held, D.H. Solina, D.T. Keane, W.J. Haag, P.M. Horn, G. Grinstein; Phys. Rev. Lett. {\bf 65}, 1120 (1990)
%
\bibitem{Oslo}
V. Frette, K. Christensen, A. Malthe - Sorensen, J. Feder, T. Jossang, P. Meakin; Nature {\bf 376}, 49 (1996)
%
\bibitem{Al1}
L.A.N. Amaral, K.B. Lauritsen; Phys. Rev. {\bf E54}, R4512 (1996)
%
\bibitem{Al2}
L.A.N. Amaral, K.B. Lauritsen; Physica A {\bf 231}, 608 (1996)
%
\bibitem{Al3}
L.A.N. Amaral, K.B. Lauritsen; Phys. Rev. {\bf E56}, 231 (1997) 
%
\bibitem{CHC}
K. Christensen, A. Corral, V. Frette, J. Feder, T. Jossang; Phys. Rev. Lett. {\bf 77}, 107 (1996)
%
\bibitem{KAD}
L. Kadanoff, S.R. Nagel, L. Wu, S.M. Zhou; Phys. Rev. {\bf A39}, 6524 (1989)
%
\bibitem{Mark}
M. Marko\v sov\'a, M.H. Jensen, K.B. Lauritsen, K. Sneppen; Phys. Rev. {\bf E}, R2085 (1997)
%
\bibitem{TDh}
B. Tadi\'c, D. Dhar; Phys. Rev. Lett. {\bf 79}, 1519 (1999)
%
\bibitem{PB}
M. Paczuski, S. Boetcher; Phys. Rev. Lett. {\bf 77}, 121 (1996)
%
\bibitem{HR}
D.A. Head, G.J. Rodges; Phys. Rev. {\bf E55}, 2573 (1997)
%
\bibitem{LSMJ}
K.B. Lauritsen, K. Sneppen, M.  Marko\v sov\'a, M.H. Jensen; Physica {\bf A247}, 1 (1997)
%
\end{thebibliography}
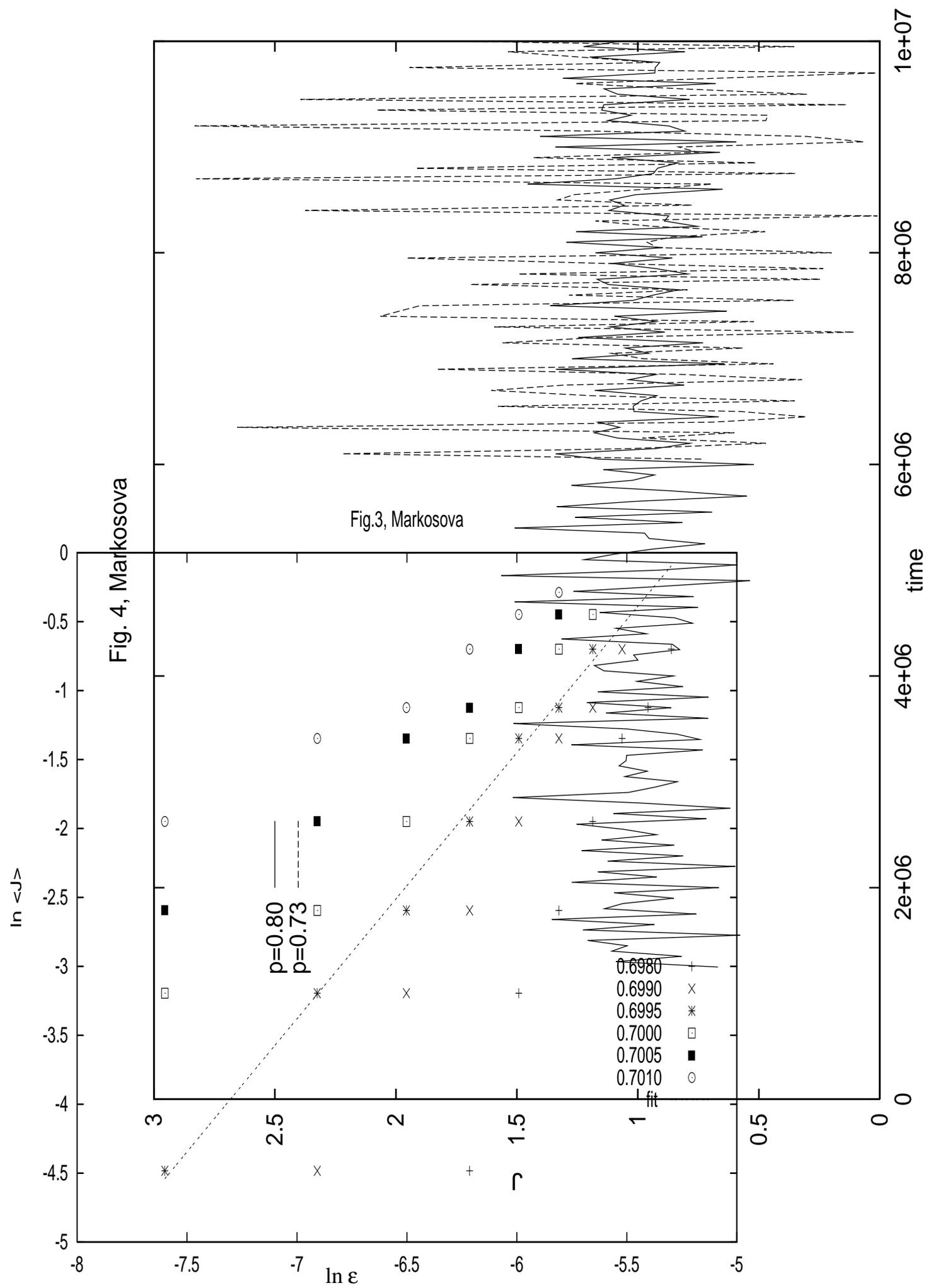
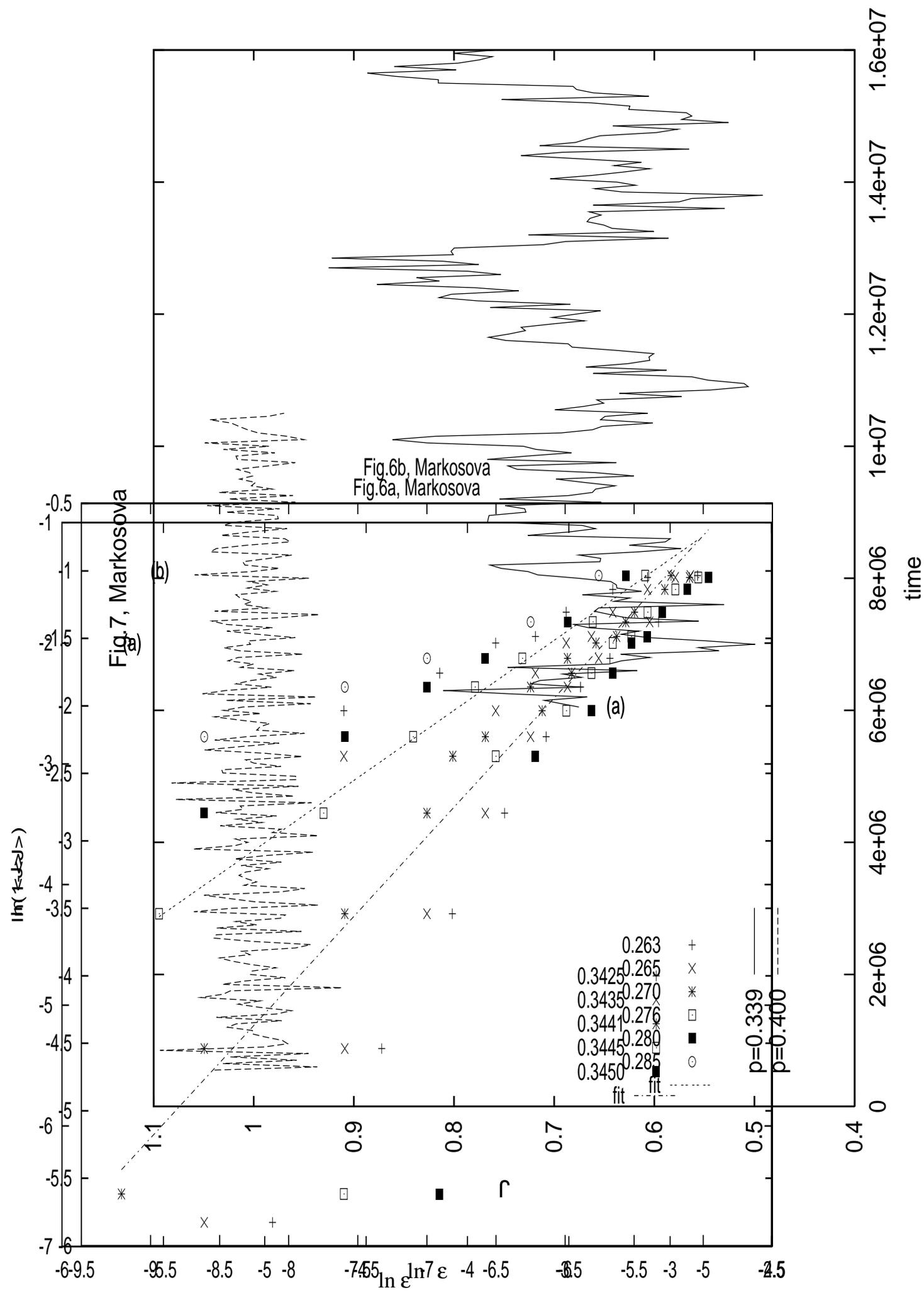
\end{document}